\documentclass[rapids]{jfm}
\usepackage{epstopdf, epsfig}
\usepackage{psfrag}
\usepackage{amssymb}
\usepackage{amsmath}
\usepackage{stmaryrd}
\usepackage{subfigure}
\usepackage{color}
\usepackage{url}
\usepackage[version=4]{mhchem}
\usepackage{caption}
\usepackage{comment}
\usepackage{rotating}
\usepackage{hhline}
\usepackage{multirow}
\graphicspath{{./Figures/}}
\def\M{{\rm{M}}}

\def\He{{\rm{He}}}

\newcommand{\f}[2]{{\frac{#1}{#2}}}

\newlength{\Oldarrayrulewidth}

\title{Compositional inhomogeneities as a source of indirect combustion noise}
\shorttitle{Compositional indirect noise}
\shortauthor{L. Magri, J. O'Brien and M. Ihme}
\author{Luca Magri, Jeff O'Brien
 \and Matthias Ihme\corresp{\email{mihme@stanford.edu}}}

\affiliation{
Center for Turbulence Research, Stanford University, \\
488 Escondido Mall, Stanford, CA 94305, United States of America
}

\begin{document}
\maketitle
\begin{abstract}
The generation of indirect combustion noise by compositional inhomogeneities is examined theoretically. For this, the compact nozzle theory of~\cite{MARBLE_CANDEL_JSV1977} is extended to a multi-component gas mixture, and  the chemical potential function is introduced as an additional acoustic source mechanism. Transfer functions for subcritical and supercritical nozzle flows are derived and the contribution of compositional noise is compared to entropy noise  and direct noise by considering an idealized nozzle downstream of the combustor exit. It is shown that compositional noise is dependent on the local mixture composition and can exceed entropy noise for fuel-lean conditions and supercritical nozzle flows.  This suggests that the compositional indirect noise requires potential consideration with the implementation of low-emission combustors.
\end{abstract}
\section{\label{SEC_INTRO}Introduction} 
The importance of engine core-noise as a relevant contributor to the overall noise emission from aircraft has been recognized, particularly for low-power engine conditions during landing and approach. Engine-core noise in aeronautical gas-turbines is commonly divided into direct and indirect noise~\citep{STRAHLE_PECS1978,CANDEL_DUROX_DUCRUIX_BIRBAUD_NOIRAY_SCHULLER_IJAA2009,DOWLING_MAHMOUDI_PCI2015}. Direct combustion noise is a source of self-noise, and describes the generation of acoustic pressure fluctuations by  unsteady heat release in the combustion chamber. In contrast, indirect combustion noise represents an induced noise-source mechanism that arises from the interaction between non-acoustic perturbations exiting the combustion chamber and downstream engine components.  The indirect noise generation by temperature inhomogeneities arising from hot and cold spots is referred to as entropy noise~\citep{CANDEL_DISS1972,MARBLE_CANDEL_JSV1977}, and indirect noise from vorticity fluctuations is referred to as vorticity noise~\citep{Cumpsty1979}. Once sound has been generated, its propagation through the engine core depends on mean flow gradients and geometric properties, which distort, diffract and reflect the acoustics. An additional noise mechanism, which is commonly neglected in the core-noise analysis, results from the modulation of the jet-noise sources by mean-flow deformations and perturbations exiting the engine core~\citep{IHME_ARFM2017}.

Direct combustion noise was examined through experimental measurements to obtain fundamental understanding about the noise-source mechanisms and the effect of fuel mixtures and operating conditions on the acoustic radiation~\citep{HURLE_PRICE_SUGDEN_THOMAS_PRSLA1968,SINGH_ZHANG_GORE_MONGEAU_FRANKEL_PCI30}, through theoretical analysis to determine the correlations for acoustic power, spectral density, and peak-frequency~\citep{RAJARAM_LIEUWEN_CST2003,CANDEL_DUROX_DUCRUIX_BIRBAUD_NOIRAY_SCHULLER_IJAA2009}, and through computational modelling using direct methods and acoustic analogies~\citep{ZHAO_FRANKEL_POF2001,IHME_PITSCH_BODONY_PCI32}.

Contributions of indirect noise to the overall core-noise emission have been examined theoretically and experimentally. These studies focused on separating the contributions to noise from the direct transmission and entropy noise. Different techniques have been employed to determine the transfer functions, including the compact nozzle theory~\citep{MARBLE_CANDEL_JSV1977}, the effective nozzle length method~\citep{STOW_DOWLING_HYNES_JFM2002,GOH_MORGANS_JSV2011}, linear nozzle element techniques~\citep{MOASE_BREAR_MANZIE_JFM2007,GIAUQUE_HUET_CLERO_JEGTP2012}, and expansion methods~\citep{DURAN_MOREAU_JFM2013}, among others. These theoretical investigations were supported by experimental studies. \cite{BAKE_RICHTER_MUEHLBAUER_KINGS_ROEHLE_THIELE_NOLL_JSV2009} conducted measurements on an entropy-wave generator to investigate entropy noise by varying the mass-flow rate, nozzle Mach number, heating power, and nozzle geometry. These investigations were extended by~\cite{KINGS_BAKE_IJSCD2010} to examine the indirect noise mechanisms arising from vorticity fluctuations. These studies showed that indirect combustion noise requires consideration in the analysis of engine-core noise and can exceed the contribution from direct noise.

Common to all of these previous theoretical and experimental investigations, however, is the restriction to a single-component gas mixture without considering effects of inhomogeneities in mixture composition on the indirect noise generation. In particular, compositional inhomogeneities can arise from incomplete mixing, air dilution, and variations in the combustor exhaust gas compositions. In this work it shown that these inhomogeneities constitute an additional indirect noise-source mechanism. The presence of compositional noise in a subsonic nozzle was first identified analytically by~\cite{IHME_ARFM2017}. The objective of this contribution is to extend this analysis by quantifying the importance of this combustion noise mechanism in subcritical and supercritical nozzles. To this end, the equations for multi-component gas mixtures are considered, and compositional fluctuations are expressed as a function of the mixture fraction. Following \cite{MARBLE_CANDEL_JSV1977}, the compact nozzle theory is used to derive transfer functions for different nozzle conditions. A parametric study is conducted to compare the relative contributions between compositional noise, entropy noise, and direct noise. 
\section{\label{SEC_NOISE_CONFINEMENT_MATH}Theoretical analysis}
The present analysis is concerned with the flow of a multi-component gas mixture through a nozzle. The following assumptions on the nozzle flow are made: 
(i) the flow is quasi one-dimensional, i.e., the variables change because of area variations but depend only on the axial coordinate; 
(ii) the gas is ideal with frozen internal energy modes so that the heat capacity only depends on the mixture composition; 
(iii) the gas is composed of $N_s$ species $Y_i$ with chemical potentials $\mu_i$; 
(iv) the flow is chemically frozen, thus, all species are expressed in terms of mixture fraction, $Z$, $Y_i=Y_i(Z)$;  
(v) perturbations have a low frequency, i.e., the Helmholtz number is small, $\He\ll1$, therefore, the flow is quasi steady and the compact nozzle assumption is valid; 
(vi) the nozzle is isentropic except across the shock for the supercritical nozzle flow.

Assumptions  (ii-iv) imply that the gas constant, $R$, is a function of the mixture fraction, because  $R=\mathcal{R}\sum_{i=1}^{N_s}Y_i(Z)/W_i$, where $W_i$ is the molar mass and $\mathcal{R}$ is the universal gas constant. 
Likewise, the specific heat capacity is a function of the mixture fraction, $c_p=\gamma/(\gamma-1)R(Z)$, and $\gamma$ is constant. The frozen-flow assumption (iv) is valid if the Damk\"{o}hler number is small, i.e., $\rm{Da}=\tau_{\textrm{flow}}/\tau_{\textrm{chem}}<1$, where $\tau_{\rm{flow}}$ is the  flow time scale and $\tau_{\rm{chem}}$ is the characteristic chemical time scale. This condition can be reformulated as $\He/(\M\,\omega\,\tau_{\rm{chem}})<1$, where $\M$ is the Mach number and $\omega$ is the perturbation frequency. For expanding nozzle flows the chemical time scale becomes large due to the reduction of the temperature and the reduced chemical reactivity of three-body recombination reactions~\citep{KECK_GILLESPIE_CF1971}. Body forces, viscous-diffusive, Soret and Dufour effects are neglected due to the nozzle flow conditions at high Reynolds numbers. 
\subsection{Governing equations}\label{sec:goveqs}
In a multi-component chemically frozen gas, the differential of the total sensible enthalpy is defined as $dh_t = dh + udu$ and the sensible enthalpy is a function of species composition and temperature, $dh=c_p dT + \sum_{i=1}^{N_s}h_idY_i$. With this and assumptions (i)-(v), the first-order perturbations of the total enthalpy, mass flow rate, and entropy read, respectively~\citep{WILLIAMS_BOOK1985} 
\begin{subeqnarray}
\f{dh_t}{h_t} &=& \f{2}{2+({\gamma-1})\M^2}\left[\f{dT}{T} + \left(\gamma-1\right)\f{\M}{c}du  +  \f{c_p'}{c_p} dZ\right],\slabel{eq:enthalpy}\\
\f{d\dot{m}}{\dot{m}} &=& \f{d\rho}{\rho} + \f{du}{\M\,c},\slabel{eq:massflowrate} \\
\f{ds}{c_p} &=& \f{dT}{T} - \f{\gamma-1}{\gamma}\f{dp}{p} + \left(\f{c_p'}{c_p} - \Psi\right) dZ, \slabel{eq:gibbs}
\end{subeqnarray}
where the symbol $'$ denotes differentiation with respect to $Z$, i.e., $c_p'=dc_p/dZ$; 
$\M=u/c$; 
$Z$ is the mixture fraction; 
$\mu_i = \mu^{0}_i+ \mathcal{R}T\ln(p_i/p_{0})$ is the chemical potential of the $i$th species~\citep{JOB_HERRMANN_EJP2006}. The superscript ``0'' denotes the standard condition, and $\Psi$=$\f{1}{c_pT}\sum^{N_s}_{i=1}\f{\mu_i}{W_i}Y_i'$ is the chemical potential function, comparing the chemical potential to the sensible enthalpy. Note that $\Psi$ is a function of the thermochemical state, $\Psi=\Psi(p,T,Z)$, however, the  dependency $(p,T,Z)$ is dropped for brevity. The chemical potential is the partial derivative of the Gibbs function, $G$, with respect to the number of moles of the $i$th species, $n_i$, at constant temperature and pressure, i.e., $\mu_i=(\partial G/\partial n_i)_{T,p,n_{j\not= i}}$. 

\begin{figure}
  \begin{center}
\includegraphics[width = 1\textwidth,clip=]{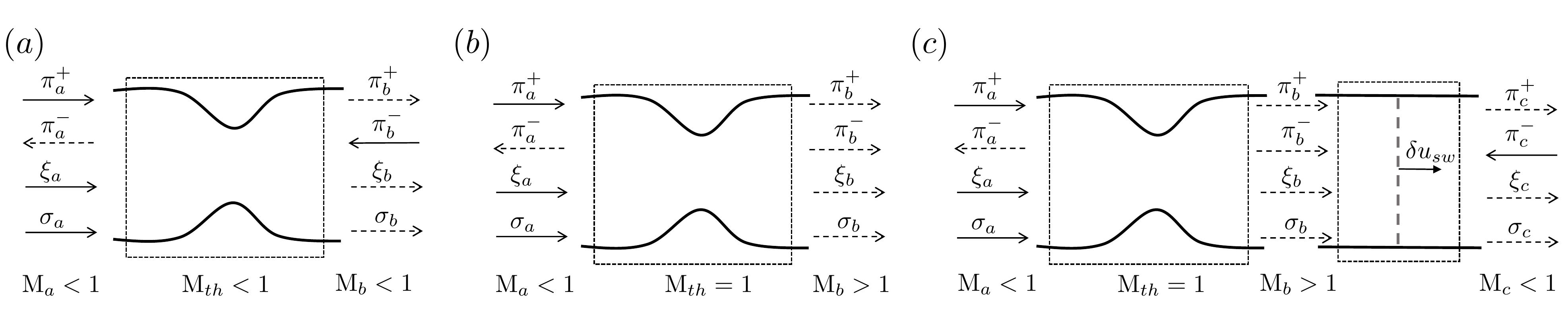}
   \caption{\label{FIG_COMPACT_NOZZLE}Acoustic, $\pi$, entropy, $\sigma$, and compositional, $\xi$, wave decomposition in a (a) subcritical nozzle, (b) supercritical nozzle and (c) supercritical nozzle with a normal shock wave. Incoming waves are denoted by solid arrows; outgoing waves are denoted by dashed arrows.  In the compact nozzle assumption, the nozzles are viewed as black boxes, depicted as dashed boxes, with properties evaluated at the inlet, $a$, and outlet, $b$. The flow region downstream of the shock wave is labeled $c$. $\M_{th}$ is the Mach number at the throat. }
   \end{center}
\end{figure}

The conservation of mass, energy, entropy and species provides the set of governing equations, which are expressed as jump conditions across the compact nozzle
\begin{equation}
 \label{EQ_COMPACT_NOZZLE_JUMP_COND}
 \llbracket d \dot{m} \rrbracket^b_a = 0,\qquad 
 \llbracket d h_{T} \rrbracket^b_a = 0,\qquad 
 \llbracket d s \rrbracket^b_a = 0,\qquad 
 \llbracket d Z \rrbracket^b_a = 0,
\end{equation}
where the indices $a$ and $b$ denote the conditions at the inlet and outlet of the nozzle, respectively (Figure~\ref{FIG_COMPACT_NOZZLE}). The system of governing equations is  closed by the differential form of the state equation $ dp/p=d\rho/\rho + (R'/R) dZ + dT/T$.  

The jump conditions \eqref{EQ_COMPACT_NOZZLE_JUMP_COND} combined with the state equation provide five relations 
for the five unknowns $dp$, $d\rho$, $dT$, $du$ and $dZ$. The entropy, $ds$, can be used as an alternative thermodynamic variable through the Gibbs relation \eqref{eq:gibbs}. 
Furthermore, $R'$ and $Y_i'$, appearing in the derivative of $c_p$ and the definition of $\Psi$,  are not state variables because they depend on the chemical composition, which will be discussed in \S\ref{Results}. 
In a choked nozzle, the variables are constrained by the condition that the mass flow rate attains a maximum, $\dot{m}^*=\sqrt{\gamma}A_{th}p_t/(\sqrt{RT_t})\left[(\gamma+1)/2\right]^{-(\gamma+1)/\left[2(\gamma-1)\right]}$, where $*$ indicates the sonic condition and the subscript ``$th$'' denotes the condition at the throat. Equating the sonic  mass flow rate to the mass flow rate \eqref{eq:massflowrate}, $d\dot{m}^*/m^*=d\dot{m}/m=0$, yields the additional condition
\begin{align}\label{eq:masscr}
&\f{ds}{2c_p}+\f{1}{2}\Psi dZ - \f{du}{\M\,c}+\f{\gamma-1}{2\gamma}\f{dp}{p}=0. 
\end{align}
The left-hand side of \eqref{eq:masscr} is equal to the perturbation Mach number, $d\M$=$du/c-\M dc/u$, where $dc=c/2\left [(R'/R)dZ+dT/T\right]$, which, therefore, is zero throughout the compact choked nozzle. 
\subsection{Nozzle transfer functions}
In the linear limit, the acoustic pressure mode is governed by a wave equation, whereas entropy and mixture-fraction modes are governed by convection equations (when the species diffusion is neglected). Extending the fundamental mode decomposition of \citet{CHU_KOVASZNAY_JFM1958} to the mixture fraction, it is inferred that these three modes are decoupled. Therefore, a characteristic decomposition can be employed. Hence, four independently evolving waves at each side of the nozzle are identified, as shown in Figure~\ref{FIG_COMPACT_NOZZLE}, which correspond to the downstream and upstream propagating acoustic waves, the convective entropy wave and the compositional waves, respectively, 
\begin{equation}\label{eq:invariants}
 \pi^\pm = \f{1}{2}\left(\f{dp}{\gamma p}\pm\f{du}{c}\right),\qquad
 \sigma = \f{ds}{c_p},\qquad \xi = dZ. 
\end{equation}

A nozzle transfer function is defined as the ratio between a single output, such as an outgoing acoustic wave, and a single input, such as an incoming entropy or compositional wave. The objective is to derive analytical transfer-function expressions for the indirect noise by compositional fluctuations, which is done by considering a subcritical nozzle, a supercritical nozzle, and a supercritical nozzle with a normal shock.
\subsubsection{Subcritical nozzle}
Using the wave decomposition \eqref{eq:invariants} and noting that $\sigma_a=\sigma_b=\sigma$ and $\xi_a=\xi_b=\xi$, the linearized equations for enthalpy and mass flow rate, \eqref{eq:enthalpy} and \eqref{eq:massflowrate}, can be written as
\begin{subeqnarray}
 \label{EQ_COMPACT_NOZZLE_FLUCT_EXPRESSIONS}
 \slabel{EQ_COMPACT_NOZZLE_FLUCT_EXPRESSIONS_H}
  \f{dh_{t}}{h_{t}} &=& \f{2(\gamma-1)}{2+{{(\gamma-1)}}\M^2}\left[(1+\M)\pi^+ + (1-\M)\pi^- + \f{\sigma + \Psi\xi}{\gamma-1} \right],\\
   \slabel{EQ_COMPACT_NOZZLE_FLUCT_EXPRESSIONS_M}
 \f{d\dot{m}}{\dot{m}} &=& \left(1+\f{1}{\M}\right)\pi^+ + \left(1-\f{1}{\M}\right)\pi^--\sigma - \Psi\xi. 
\end{subeqnarray}
Substituting equations (\ref{EQ_COMPACT_NOZZLE_FLUCT_EXPRESSIONS}) into (\ref{EQ_COMPACT_NOZZLE_JUMP_COND}) provides a linear set of algebraic equations that relates the four incoming waves ($\pi^+_a, \sigma_a, \xi_a, \pi^-_b$) to the four outgoing waves ($\pi^-_a, \sigma_b, \xi_b, \pi^+_b$). After algebraic manipulations, the transfer functions between the acoustic wave leaving the nozzle and the acoustic, entropy, and compositional waves entering the nozzle can be derived, and the resulting expressions are presented in Table~\ref{tab:tra_subch}. 
\subsubsection{Supercritical nozzle}\label{sec:chokedflow}
For a supercritical nozzle flow (Figure~\ref{FIG_COMPACT_NOZZLE}(b)), the inputs are three incoming waves ($\pi^+_a, \sigma_a, \xi_a$) and the outputs are five outgoing waves ($\pi^-_a, \sigma_b, \xi_b, \pi^+_b, \pi^-_b$). Similar to the subcritical nozzle, $\sigma_a=\sigma_b=\sigma$ and $\xi_a=\xi_b=\xi$ due to the jump conditions \eqref{EQ_COMPACT_NOZZLE_JUMP_COND}. 
As explained in \S\ref{sec:goveqs}, the perturbation of the mass flow rate provides two conditions 
\begin{subeqnarray} \label{eq:chokedconditions}
&\sigma + \Psi_a\xi + \left[\f{\M_a(\gamma-1)-2}{\M_a}\right]\pi^+_a +  \left[\f{\M_a(\gamma-1)+2}{\M_a}\right]\pi^-_a=0, \\
&\sigma + \Psi_b\xi + \left[\f{\M_b(\gamma-1)-2}{\M_b}\right]\pi^+_b +  \left[\f{\M_b(\gamma-1)+2}{\M_b}\right]\pi^-_b=0. 
\end{subeqnarray}
By relating outgoing and ingoing waves, the transfer functions are obtained and are summarized in Table~\ref{tab:tra_subch}. The relations for the outgoing wave $\pi^-_b$, needed in \S\ref{sec:sw}, can be derived by antisymmetry, substituting $\M_b\rightarrow-\M_b$ into the transfer functions for $\pi^+_b$. 
\subsubsection{Supercritical nozzle with shock wave}\label{sec:sw}
It is assumed that the pressure at the nozzle exit is such that a shock wave occurs downstream of the choked condition. The disturbances ($\pi^+_b$, $\pi^-_b$, $\sigma$, $\xi$) impinge on the shock wave and move its position by a first-order perturbation to the velocity, $\delta u_{sw}$  (Figure~\ref{FIG_COMPACT_NOZZLE}(c)). To calculate the outgoing waves downstream of the shock wave ($\pi_c^+$, $\sigma_c$), noting that $\xi_c=\xi$ because the flow is frozen, the following procedure is implemented \citep{MARBLE_CANDEL_JSV1977}.   
First, the flow variables are expressed in the reference frame attached to the shock wave by a Galilean transformation $u_b-\delta u_{sw}$ and $u_c-\delta u_{sw}$, which defines the effective Mach number $\M_{b,sw}=\M_b(1-\delta u_{sw}/u_b)$. Secondly, the velocity, pressure and density Rankine-Hugoniot relations are linearized by considering the linearized effective Mach number, $d\M_{b,sw}=d\M_b-\M_b\delta u_{sw}/u_b$, where $d\M_b=0$  for a choked flow (see \eqref{eq:masscr}) and $d\M_b\delta u_{sw}$ is neglected because it is of  higher order. These read, respectively 
\begin{subeqnarray}\label{eq:linrh}
\f{du_c}{u_c}-\f{du_b}{u_b} &=& \f{u_b}{u_c}\left[\f{4}{\left(\gamma+1\right)\M_b^2} - \f{u_c}{u_b} +1\right] \f{\delta u_{sw}}{u_b}, \\
\f{dp_c}{\gamma p_c} - \f{dp_b}{\gamma p_b} &=&-\f{p_b}{p_c} \f{4\M_b^2}{\left(\gamma+1\right)} \f{\delta u_{sw}}{u_b},\\
\f{d\rho_c}{\rho_c} - \f{d\rho_b}{\rho_b} &=&-\f{\rho_b}{\rho_c} \f{\left(\gamma+1\right)\M_b^2}{\left[1+\f{1}{2}\left(\gamma-1\right)\M_b^2\right]^2} \f{\delta u_{sw}}{u_b}. 
\end{subeqnarray}
Equations~\eqref{eq:linrh} along with the linearized continuity equation, provide the perturbed flow state after the shock wave ($du_c$, $dp_c$, $d\rho_c$) and the shock-wave velocity, $\delta u_{sw}$. (Note that $\delta u_{sw}$ does not depend on frequency because of the compact-nozzle assumption.)
Finally, the wave decomposition \eqref{eq:invariants} and Gibbs relation for the density, $d\rho/\rho=dp / (\gamma p) -ds/c_p-\Psi dZ$ are applied. With this, the outgoing acoustic and entropy waves after the shock read, respectively 
\begin{subeqnarray}
\slabel{eq:pic}
\pi^+_c&=& \left( \frac{1+2\M_c^2\M_b+\M_b^2}{1+2\M_b^2\M_c+\M_b^2}\right){\pi_b^+}+  \left(\frac{1-2\M_c^2\M_b+\M_b^2}{1+2\M_b^2\M_c+\M_b^2}\right){\pi_b^-}, \\ 
\slabel{eq:sic}
 \sigma_c&=&\sigma-(\Psi_c-\Psi_b)\xi +\left[\frac{(\gamma-1)(\M_b^2-1)^2}{\M_b^2(2+(\gamma-1)\M_b^2)}\right]\left(\pi^+_c+\pi^-_c-\pi^+_b-\pi^-_b\right). 
\end{subeqnarray}
Note that $u_c$, $p_c$ and $\rho_c$ in~\eqref{eq:linrh} depend on the Mach number downstream of the shock wave, $\M_c$, which, in turn, is related to the Mach number upstream of the shock wave, $\M_b$, through the normal shock wave relation
\begin{equation}
\M^2_c={\frac{\M_b^2(\gamma-1)+2}{2\gamma\M_b^2-(\gamma-1)}}.
\end{equation} 

The transfer functions $\pi^+_c/\pi^+_a$, $\pi^+_c/\sigma_a$ and $\pi^+_c/\xi_a$ can be derived by substituting  the transfer functions for $\pi^+_b$ and $\pi^-_b$ for a choked nozzle into \eqref{eq:pic}. It is interesting to note that the difference in the chemical potential function in \eqref{eq:sic}, $(\Psi_c-\Psi_b)\xi$, is a further source of entropy across the shock wave. 
\begin{table}
\centering
\renewcommand{\arraystretch}{2} 
\begin{tabular}{l | c | c  }
  & Subcritical nozzle & Supercritical  nozzle       \\ \hhline{===}
  $\pi^+_b/\pi^+_a$ &
  $\f{2(1+\M_a)\M_b}{(1+\M_b)(\M_a+\M_b)}\f{\left[2+{(\gamma-1)}\M_b^2\right]}{\left[2+{(\gamma-1)}\M_a\M_b\right]} $ &
  $\f{2+(\gamma-1)\M_b}{2+(\gamma-1)\M_a}$\\ \cline{1-3}
  $\pi^+_b/\sigma_a$ & 
  $\f{(\M_b-\M_a)\M_b}{(1+\M_b)\left[2+{(\gamma-1)}\M_a\M_b\right]}$ & 
  $\f{1}{2}\f{\M_b - \M_a}{2+(\gamma-1)\M_a}$ \\\cline{1-3}
  $\pi^+_b/\xi_a$      &
  \begin{tabular}{@{}c@{}}
  $\f{(\gamma-1)(\Psi_b-\Psi_a)\left[2+{(\gamma-1)}\M_b^2\right]\M_a\M_b}{(\gamma-1)(1+\M_b)(\M_a+\M_b)\left[2+{(\gamma-1)}\M_a\M_b\right]}$\\
  $+\f{\M_b\left[2(\Psi_a-\Psi_b)+{(\gamma-1)}(\Psi_a\M_b^2-\Psi_b\M_a^2)\right]}{(\gamma-1)(1+\M_b)(\M_a+\M_b)\left[2+{(\gamma-1)}\M_a\M_b\right]}$\end{tabular} &
  $\f{1}{2(\gamma-1)}\left[-\Psi_b + \f{2+(\gamma-1)\M_b}{2+(\gamma-1)\M_a}\Psi_a\right]$
\end{tabular}
\caption{Transfer functions for subcritical and supercritical nozzles. The transfer functions $\pi^+_b/\pi^+_a$ and $\pi^+_b/\sigma_a$  were derived by~\cite{MARBLE_CANDEL_JSV1977}. }\label{tab:tra_subch} 
\end{table}
\subsection{Comments on the theoretical analysis}
The indirect noise generated by compositional inhomogeneities is physically due to the transfer of chemical potential energy into acoustic energy through the accelerating flow.  
From the transfer functions and transmission relations that were derived in the previous sections, the following limits are worth considering.  
First, when $\M_a\rightarrow \M_b$, the compositional noise tends to zero because $\Psi_a\rightarrow\Psi_b$ (Table~\ref{tab:tra_subch}). Secondly, in the limit of a constant chemical potential function, $\Psi_a = \Psi_b$, it can be shown that the ratio between compositional noise and entropy noise for all the three cases tends to the same limit $\Psi$. This limit physically signifies that the indirect noise ratio is independent of the flow conditions, and is only a function of the thermodynamic and compositional state at the inlet, under the compact-nozzle assumption. 
\section{\label{Results}Results}
The analysis developed in the previous section is applied to a flow-path configuration to quantify the relative contribution of compositional noise to the overall combustion noise. For this, we consider an idealized configuration in which the combustor exhaust-gas composition enters the nozzle. This exhaust-gas composition is represented by the solution of a series of one-dimensional strained diffusion flames~\citep{PETERS_BOOK2000} that include the equilibrium composition, typically observed at low-power cruise conditions, and highly strained combustion conditions representative of high-load operation. The flame solutions are generated by considering $n$-dodecane (\ce{C12H26}), a kerosene surrogate, as fuel and air in the oxidizer stream at operating conditions of 295\,K and ambient pressure. The flame structure is parameterized by the mixture fraction, with $Z = 0$ corresponding to the oxidizer stream and $Z = 1$ corresponding to the pure fuel stream. The flame structure is obtained from the steady-state solution of the conservation equations for continuity, species, and energy, which are solved using the {\sc{Cantera}} software package~\citep{CANTERA}. The reaction chemistry is described by a 24-species mechanism~\citep{Vie15}, which provides an accurate flame representation at these conditions. 

\begin{figure}
 \centering
\includegraphics[width=0.8\textwidth]{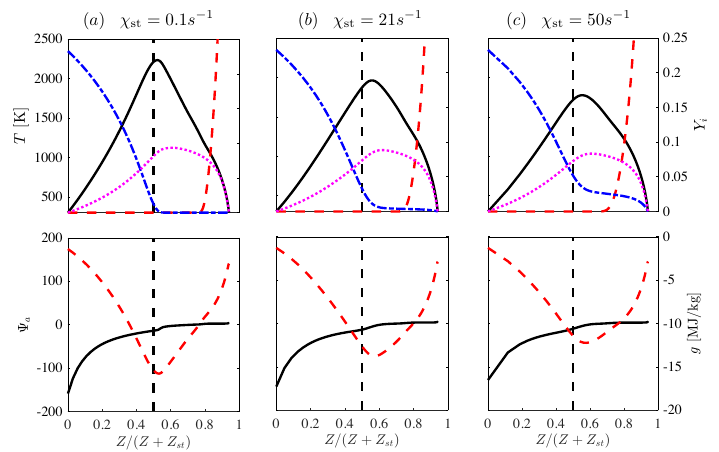}
 \caption{\label{FIG_FLAMELET_STRUCTURE}Representation of one-dimensional diffusion flame in mixture-fraction composition space for three different scalar dissipation rates. First row: flame structure, showing  temperature $T$ (solid black lines),  oxygen mass fraction $Y_{\ce{O2}}$ (blue dot-dashed lines),  $n$-dodecane mass fraction $Y_{\ce{C12H26}}$ (red dashed lines), and water mass fraction $Y_{\ce{H2O}}$ (magenta dotted lines). Second row: chemical potential function $\Psi_a$ (solid black lines) and specific Gibbs energy of the mixture, $g = \sum_i \frac{\mu_i}{W_i}Y_i$ (dashed red lines). $Z_{\rm{st}}$ is the stoichiometric mixture fraction. Operating conditions: \ce{C12H26}/air combustion, $T_{\rm{fuel}}=T_{\rm{ox}} = 295\,\rm{K}$, $p=1\,\rm{bar}$.}
\end{figure}

The degree of straining, i.e., the deviation from equilibrium, is characterized by the scalar dissipation rate, $\chi = 2 \alpha |\nabla Z|^2$, where $\alpha$ is the diffusivity of the mixture fraction, and $\chi$ is evaluated at the stoichiometric condition, corresponding to a value of $Z_{\rm{st}}=0.063.$ Large values of $\chi_{\rm{st}}$ correspond to high strain rates conditions, in which diffusive transport of heat away from the flame exceeds the heat-release. Flame extinction occurs when $\chi_{\rm{st}}$ exceeds the quenching limit. The present study considers three physically significant operating conditions with (a) $\chi_{\rm{st}}=0.1\,$s$^{-1}$ (quasi unstrained condition near equilibrium), (b) $\chi_{\rm{st}}=21\,$s$^{-1}$ (intermediately strained flame condition), and (c) $\chi_{\rm{st}}=50\,$s$^{-1}$ (highly strained flame at condition near extinction). The structure of each flame together with the chemical potential function and the specific Gibbs energy is shown in Figure~\ref{FIG_FLAMELET_STRUCTURE}. The results are presented as a function of the transformed mixture-fraction coordinate $Z/(Z + Z_{\rm{st}})$, which divides the plot evenly between lean ($Z < Z_{\rm{st}}$) and rich ($Z > Z_{\rm{st}}$) conditions. 

These flame solutions can be interpreted as an idealized representation of the gas composition exiting the combustor. The combustor operates at a global equivalence ratio $\phi$ corresponding to a mean mixture fraction $Z$, with $Z=\phi Z_{\rm{st}}/[Z_{\rm{st}}(\phi-1)+1]$~\citep{PETERS_BOOK2000}, and the corresponding thermochemical state is then taken from the flame solution of Figure~\ref{FIG_FLAMELET_STRUCTURE}. In addition to temperature fluctuations, which are known to generate entropy noise, incomplete mixing, turbulence, and other unsteady effects give rise to fluctuations in $Z$. This has the potential to produce compositional noise downstream of the combustor. To assess the compositional noise that is generated, the combustor exhaust composition for a given value of $Z$ is isentropically compressed through an ideal nozzle, keeping the mean mixture composition frozen at this flame state. Since gas turbine combustors typically operate at subsonic conditions, without loss of generality, it is assumed that $\M_a = 0$. The transfer functions of Table~\ref{tab:tra_subch} and the shock-wave case (\S\ref{sec:sw}) are then evaluated over the full mixture-fraction space and a range of relevant nozzle-exit Mach numbers. 

\begin{figure}
  \centering
  \begin{tabular}{ c c  c  c}
  & $\chi_{\rm{st}} = 0.1\,\text{s}^{-1}$& $\chi_{\rm{st}} = 21\,\text{s}^{-1}$ & $\chi_{\rm{st}} = 50\,\text{s}^{-1}$ \\ \hline\hline
  \multicolumn{4} {c} {Subcritical and supercritical nozzle}\\\hline
  \begin{sideways}\hspace*{3mm} $\log_{10}\Big\vert \f{\pi_b^+\slash\xi}{\pi_b^+\slash\pi_a^+} \Big\vert$ \end{sideways}
  & \includegraphics[height=.15\textheight]{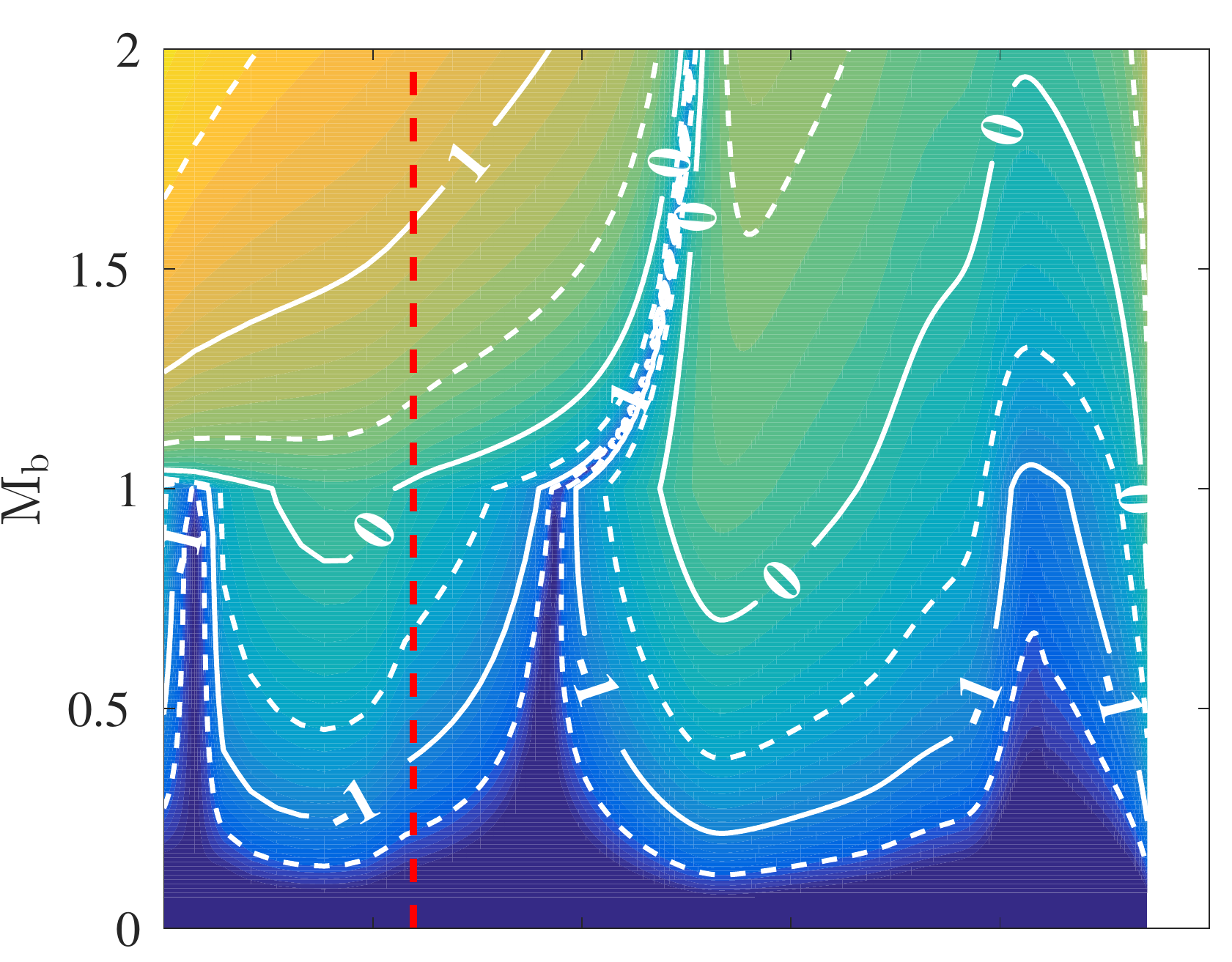}
  & \includegraphics[height=.15\textheight]{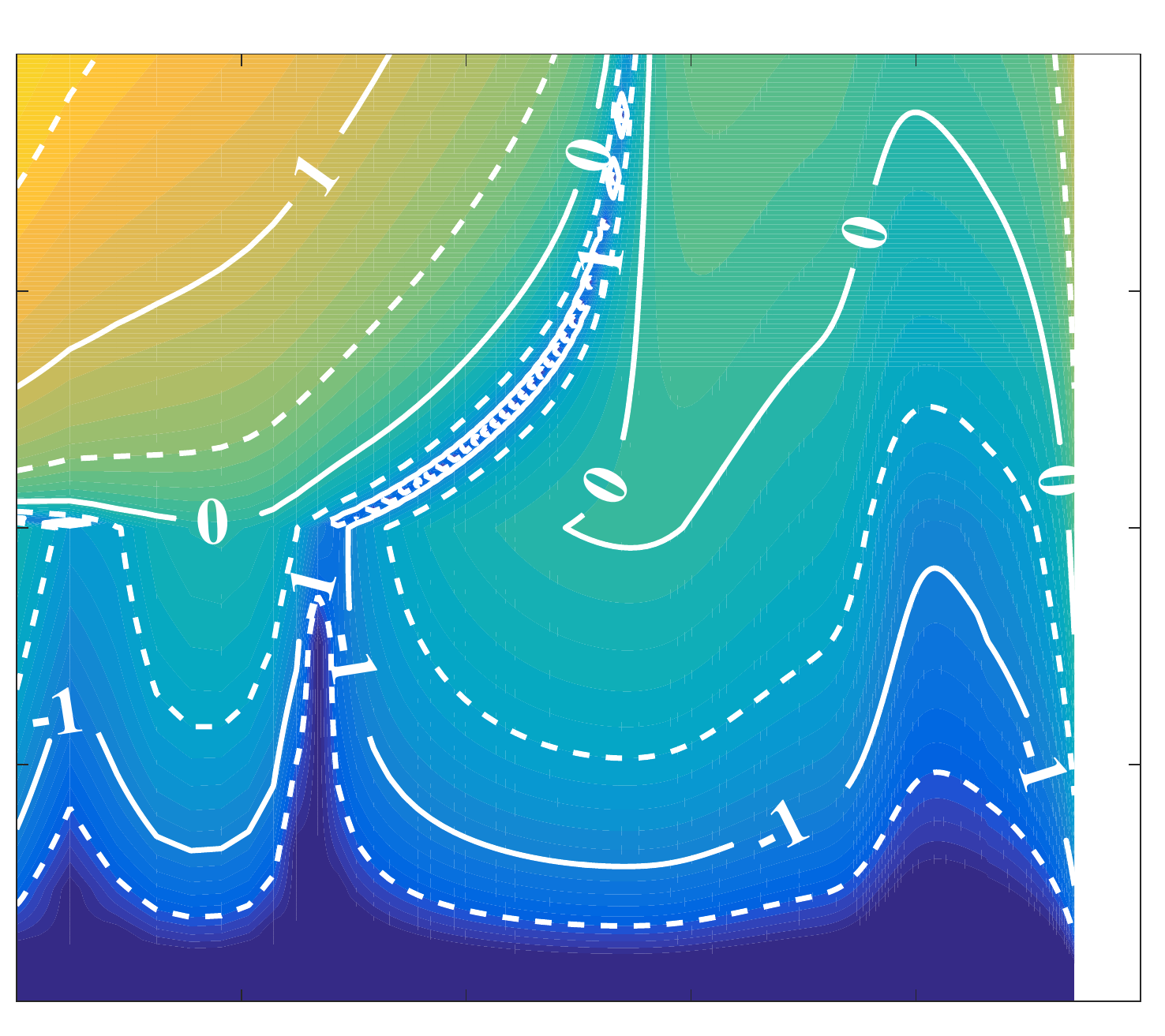}
  & \includegraphics[height=.15\textheight]{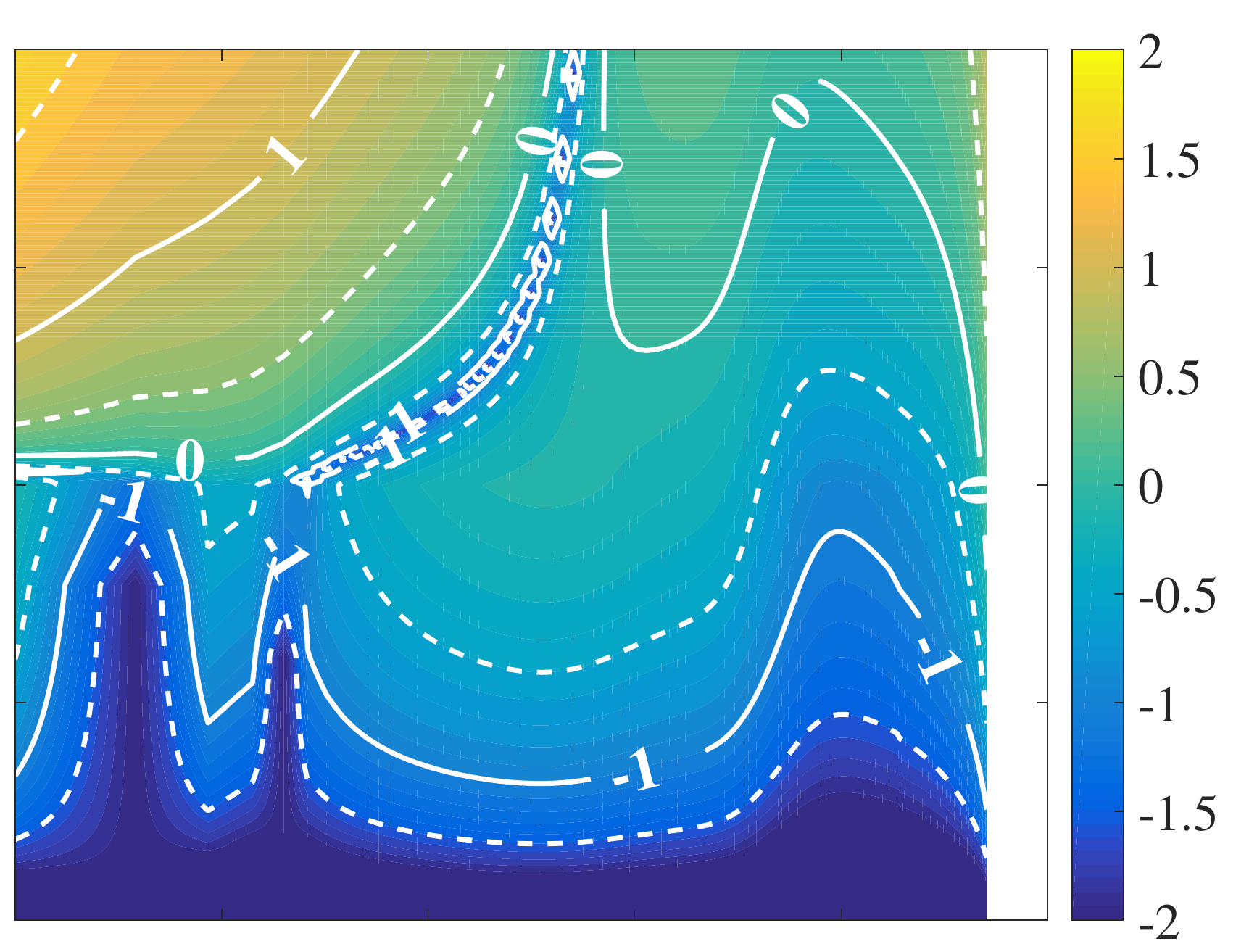}
   \\ 
  \begin{sideways}\hspace*{5mm} $\log_{10}\Big\vert \f{\pi_b^+\slash\xi_a}{\pi_b^+\slash\sigma_a} \Big\vert$\end{sideways}
  & \includegraphics[height=.15\textheight]{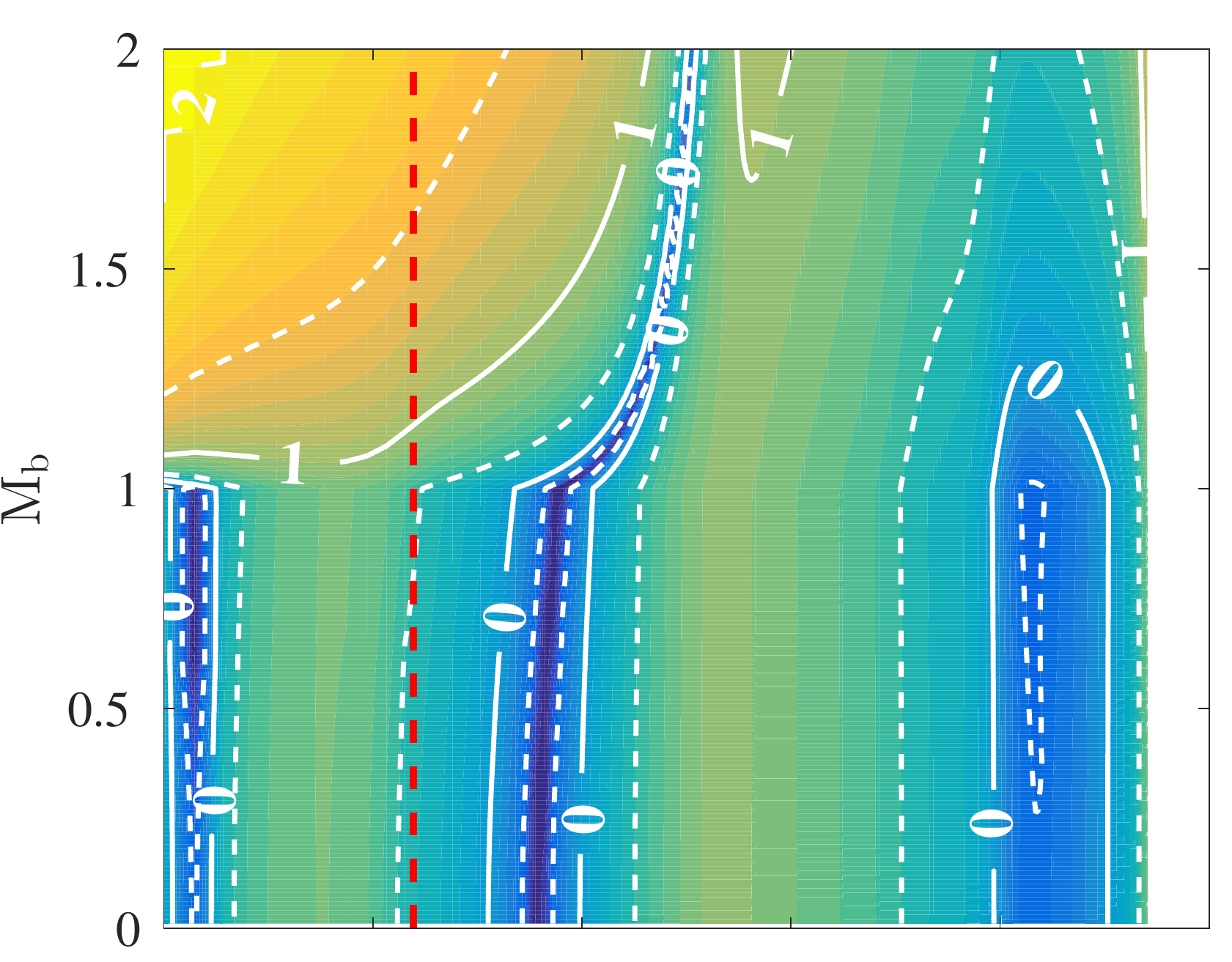}
  & \includegraphics[height=.15\textheight]{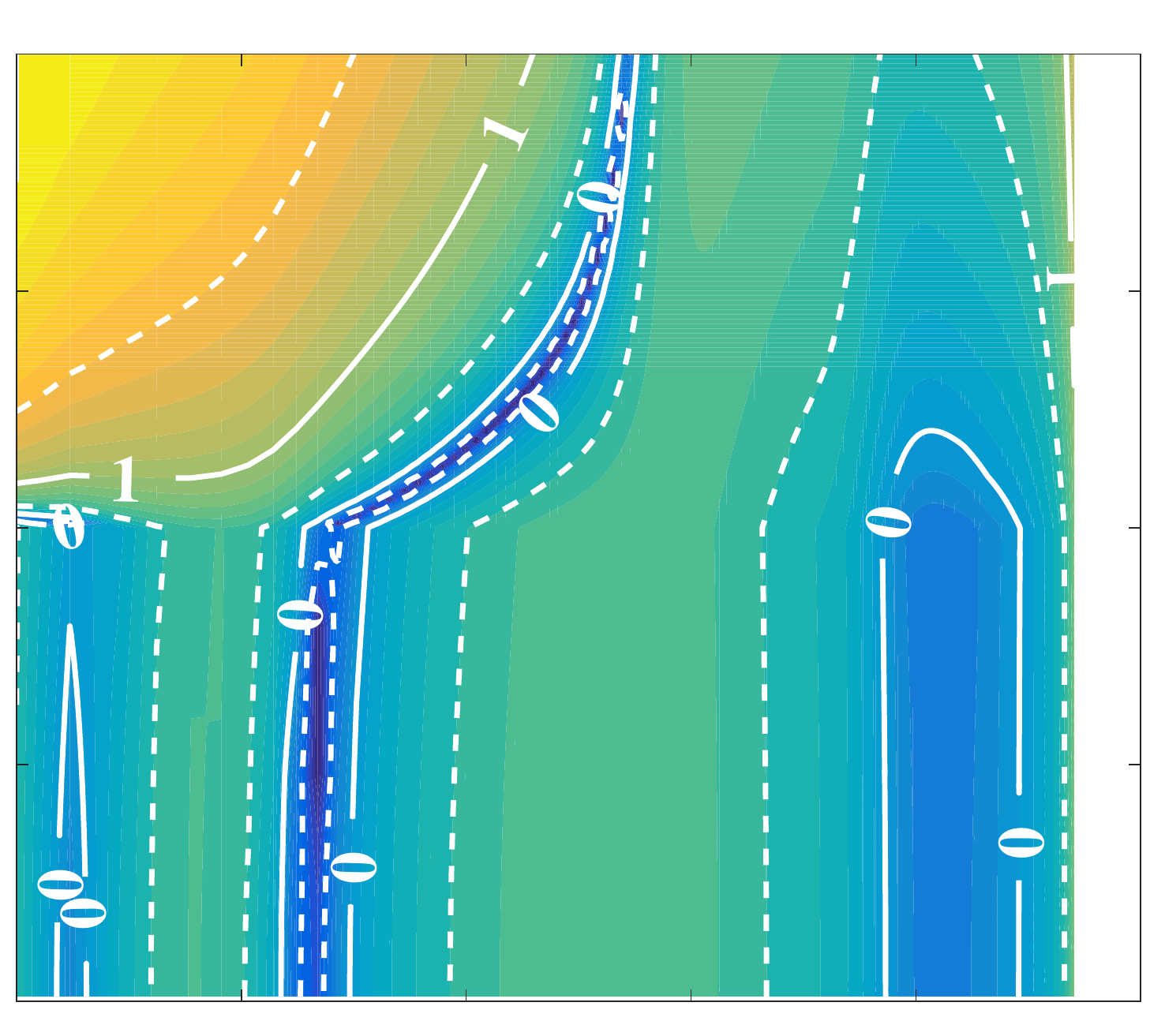}
  & \includegraphics[height=.15\textheight]{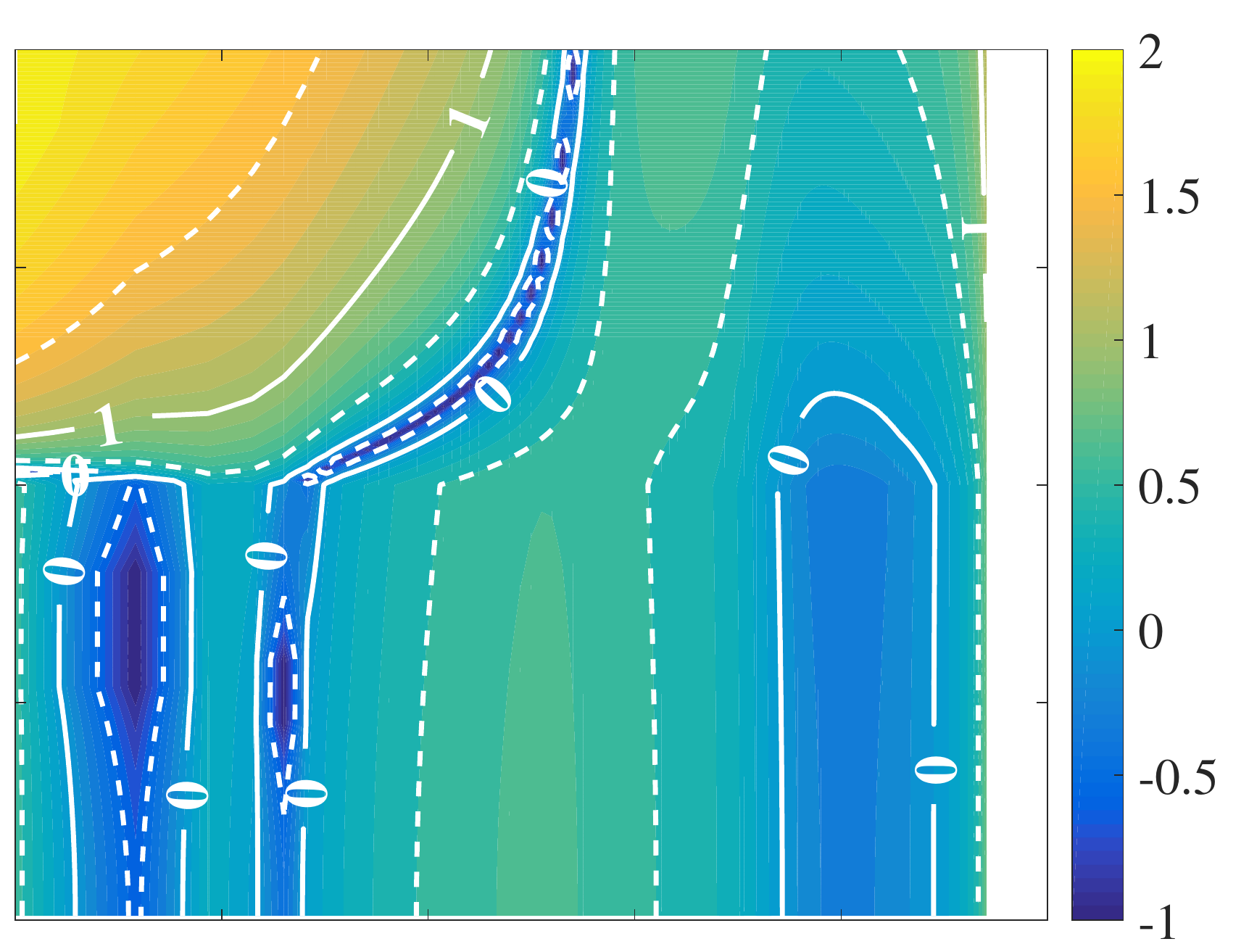}
 \\ 
 \hline
  \multicolumn{4} {c} {Supercritical nozzle with shock wave}\\\hline
    \begin{sideways}\hspace*{3mm} $\log_{10}\Big\vert \f{\pi_c^+\slash\xi_a}{\pi_c^+\slash\pi_a^+} \Big\vert$\end{sideways}
  & \includegraphics[height=.15\textheight]{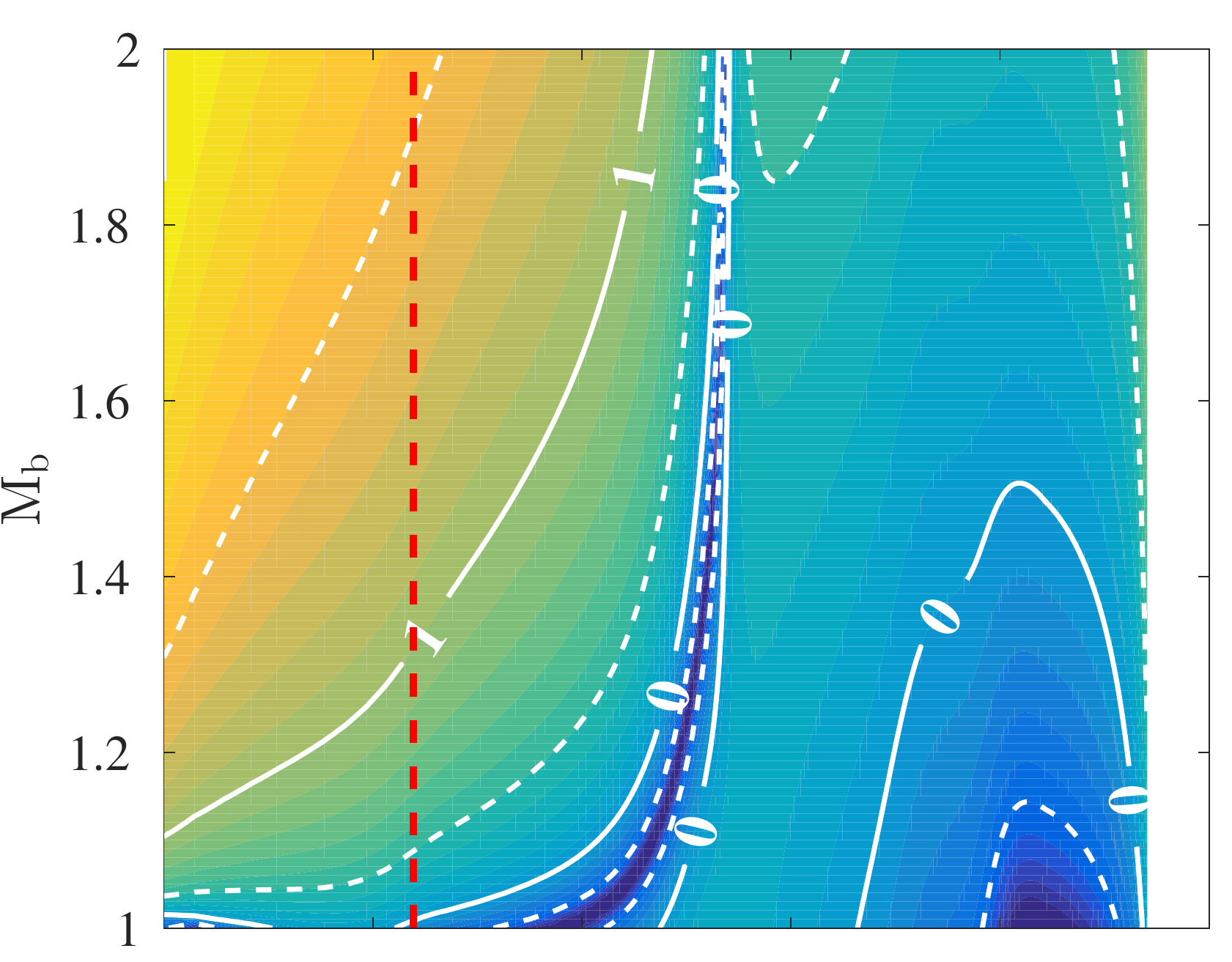}
  & \includegraphics[height=.15\textheight]{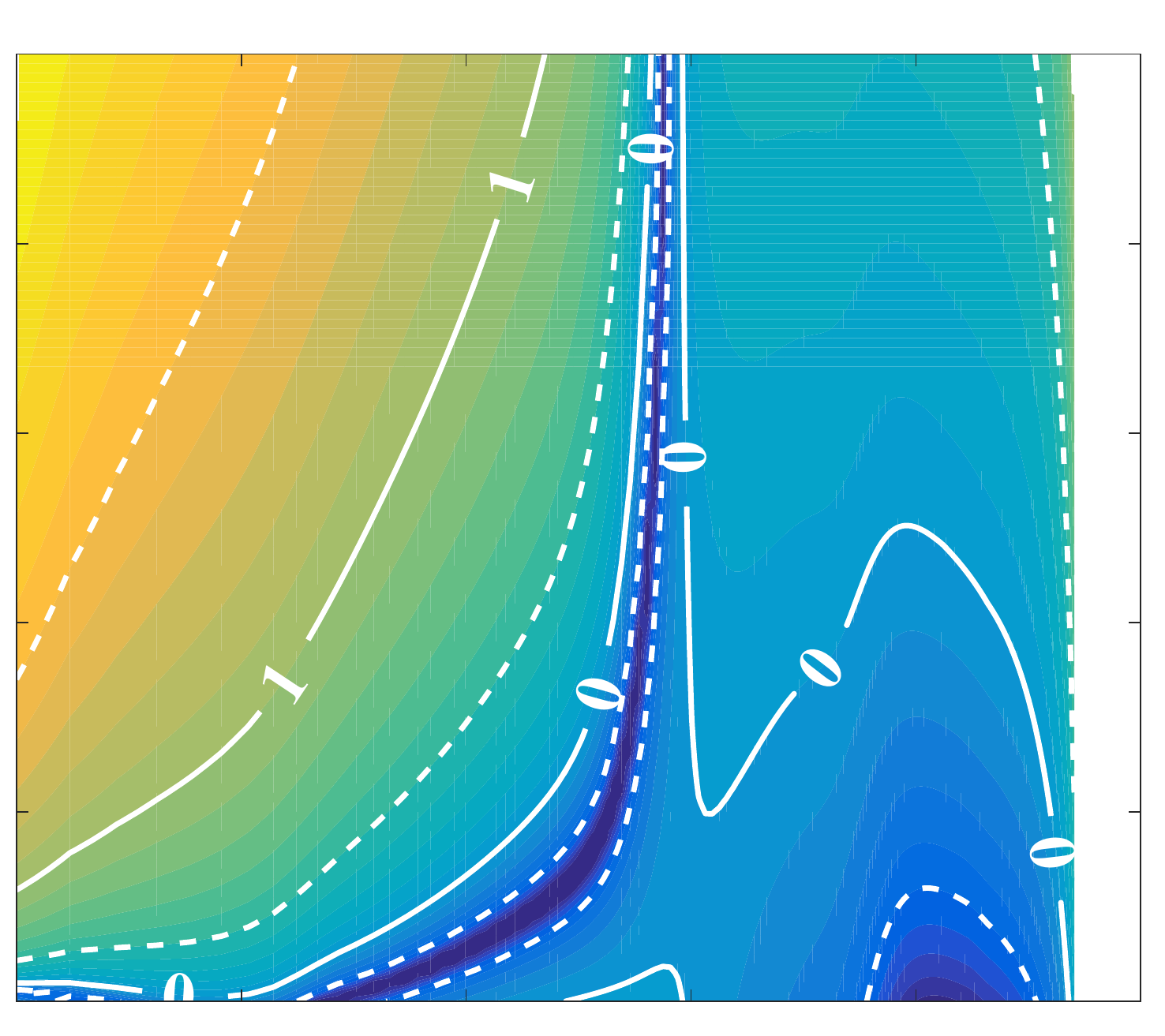}
  & \includegraphics[height=.15\textheight]{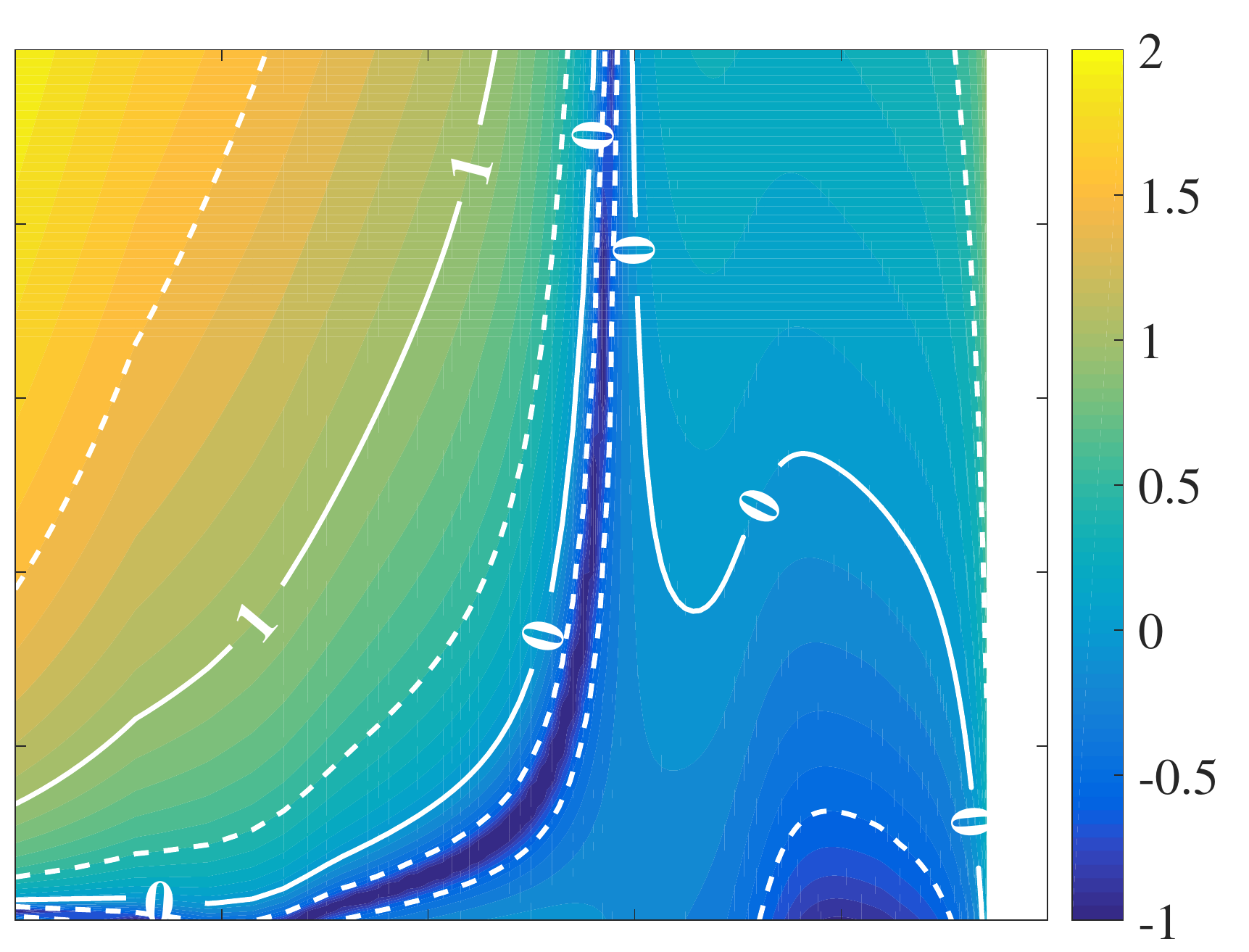}
  \\ 
      \begin{sideways} \hspace*{5mm} $\log_{10}\Big\vert \f{\pi_c^+\slash\xi_a}{\pi_c^+\slash\sigma_a} \Big\vert$\end{sideways}
  & \includegraphics[height=.17\textheight]{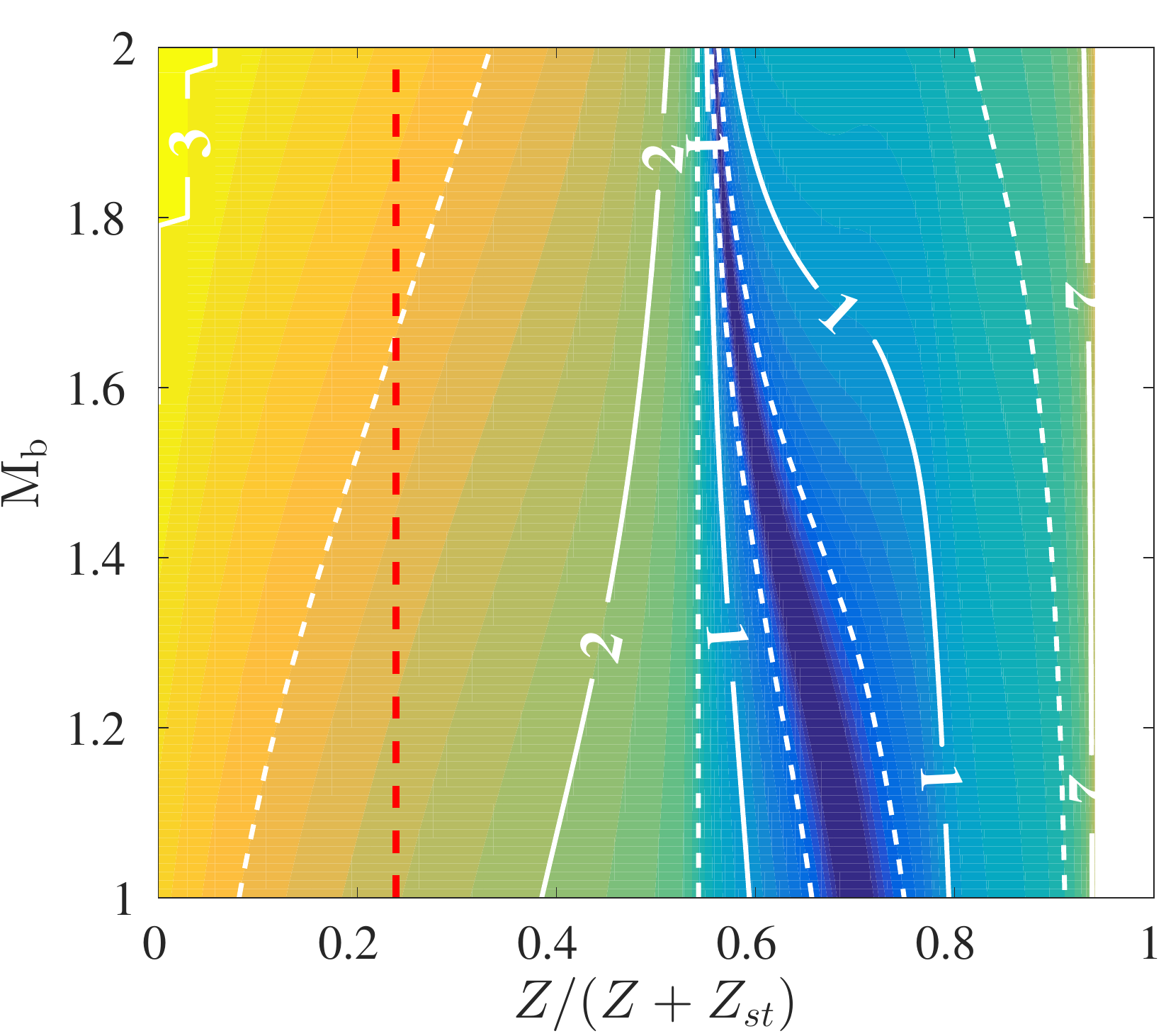}
  & \includegraphics[height=.17\textheight]{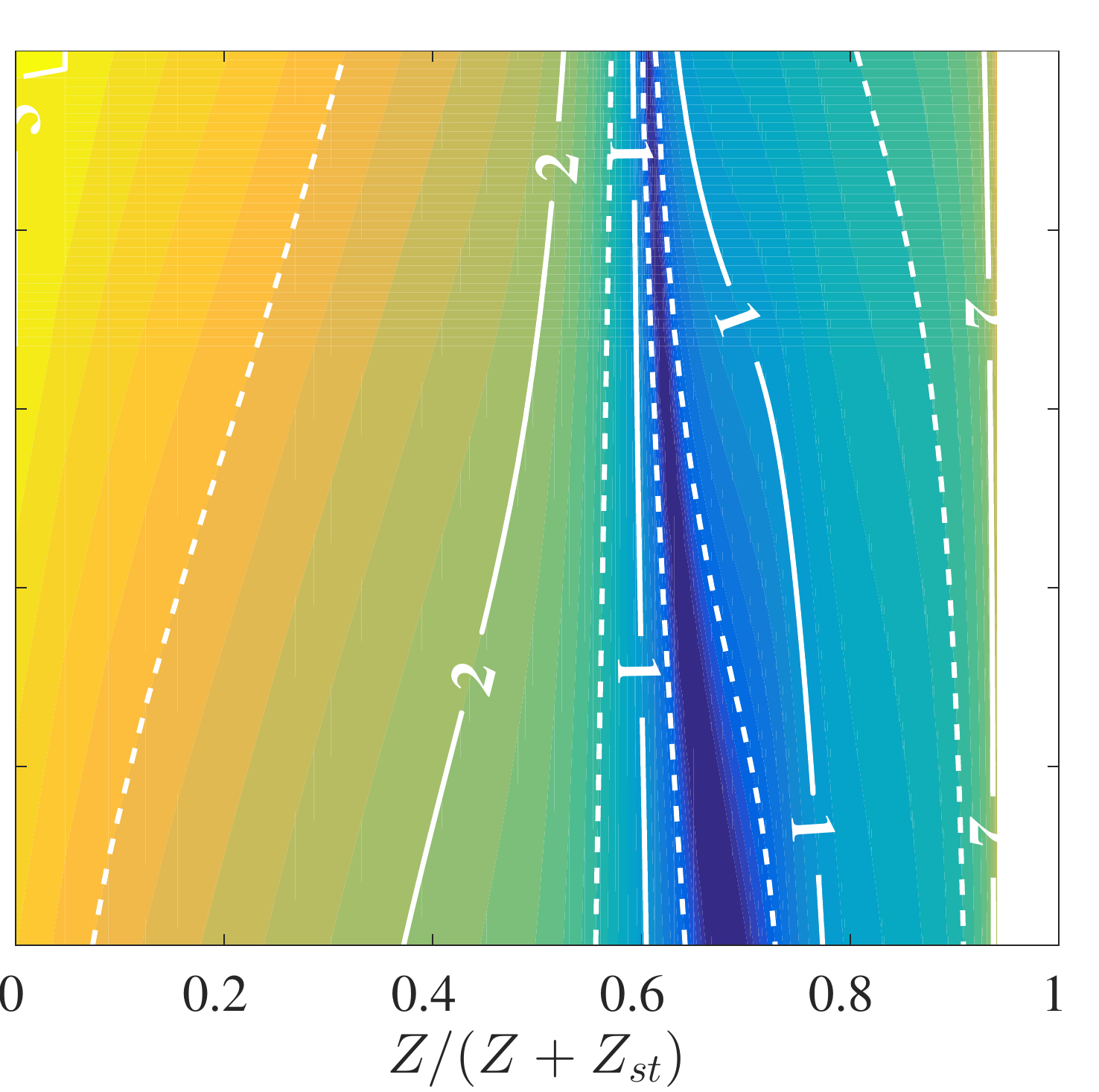}
  & \includegraphics[height=.17\textheight]{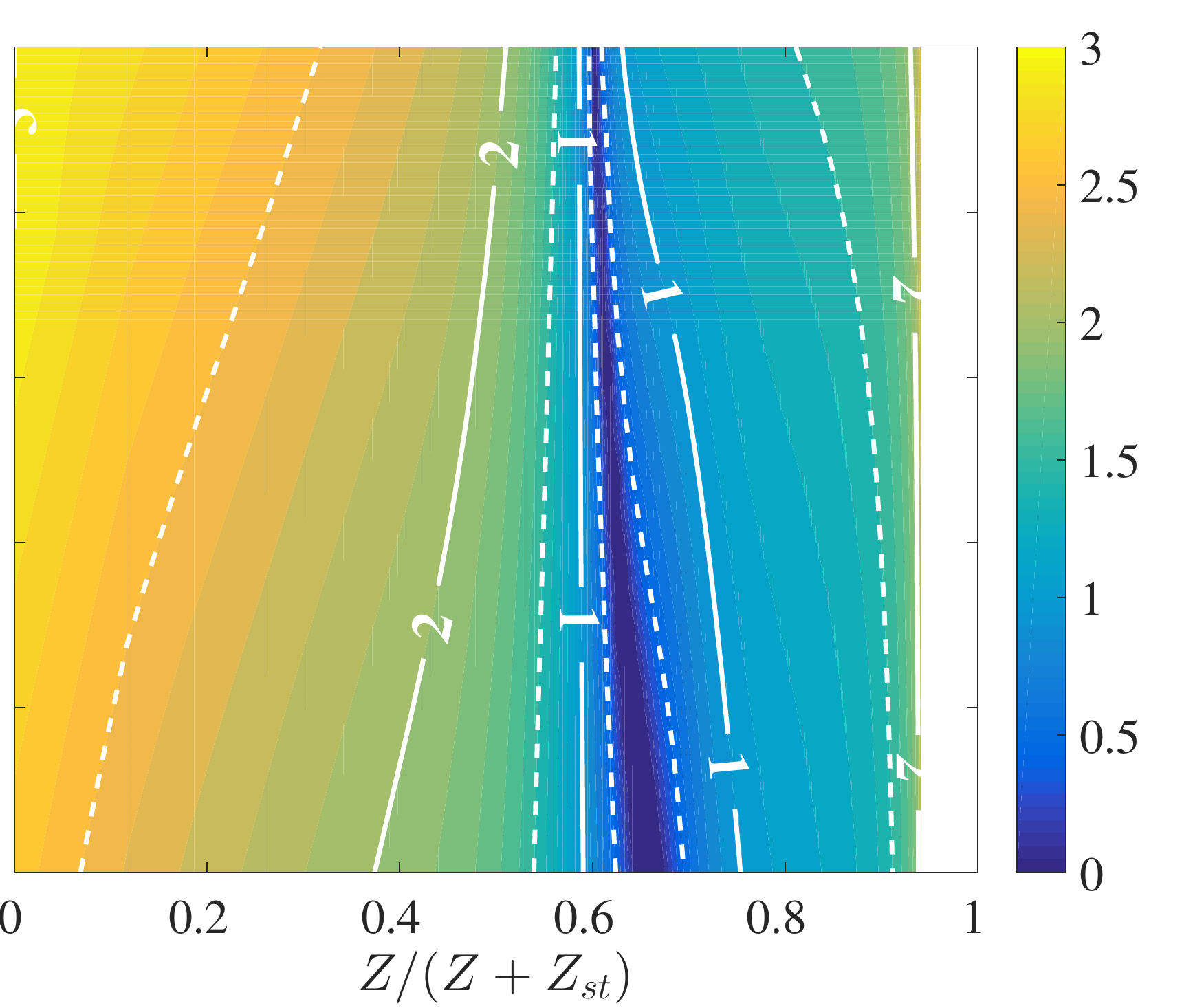}
  \end{tabular}
\caption{\label{FIG_TRANSFER_PLOTS}Transfer function ratios for (first row) compositional noise to direct noise; (second row) compositional noise to entropy noise; (third row) compositional noise to direct noise after a shock wave; and (fourth row) compositional noise to entropy noise after a shock wave. The columns correspond to the three combustor exit conditions of Figure~\ref{FIG_FLAMELET_STRUCTURE}. The vertical red dashed line indicates the condition of equivalence ratio of $\phi=0.3$.}
 \end{figure}

The transfer function ratios between compositional noise, direct noise, and entropy noise for different nozzle flows and combustor exhaust compositions are presented in Figure~\ref{FIG_TRANSFER_PLOTS}. The first and second rows of Figure~\ref{FIG_TRANSFER_PLOTS}, show the ratio of the transfer functions between compositional and direct noise and between compositional and entropy noise, respectively, for an ideally expanded nozzle. The third and fourth rows show the corresponding results for the nozzle flow with shock wave. 

From these results, it can be seen that the transfer function for the compositional noise depends on nozzle-exit condition, gas composition and dissipation rate. This is most pronounced for fuel-lean and supersonic conditions. The dependence of the compositional noise on the gas mixture at fuel-lean conditions is particularly noteworthy because it corresponds to the typical operating regime of modern gas-turbine engines. This sensitivity is a direct result of stronger variations of the mixture composition and inherent differences in the chemical potential at fuel-lean conditions. This suggests that variations in the equivalence ratio, for instance during the engine operation or the consideration of low-emission combustor concepts, can lead to noise modulation by induced compositional noise, in addition to direct and entropy noise. 

Effects of increasing scalar dissipation rate are mainly evident for fuel-lean conditions, which is attributed to the leakage of reactants and incomplete combustion thereby reducing the chemical potential function. Figure~\ref{FIG_FLAMELET_STRUCTURE} shows the variation of the chemical potential function as well as the specific Gibbs energy as a function of the  mixture fraction. It can be seen that the variation in $g$ and correspondingly the magnitude of $\Psi$ are largest at fuel-lean conditions. While this broadening effect is most easily seen in physical space, it also has a weaker sensitivity in mixture fraction space, leading to the differences with respect to $\chi_{\rm{st}}$. Simulations at higher pressure conditions show that the pressure has a negligible effect on the magnitude of the transfer function (results not shown). This is likely because differences in the chemical potential vary weakly with the  temperature and pressure because the chemical composition of the flame is not strongly changing along these paths, and the chemical potential depends logarithmically on the pressure. At extreme temperatures and pressures, where dissociation of diatomic gases occurs, the sensitivity to the thermodynamic state is likely to be much stronger.

To connect these results to practical applications, we provide an estimate of the ratio of composition noise to entropy noise by multiplying the corresponding transfer function ratios with the factor $\xi_a/\sigma_a=\delta Z_a/(\delta T_a/T_a)$. This factor is estimated by considering that the mixture composition at the combustor exit reaches equilibrium with a mean temperature of $T_a=1085$ K, corresponding to an equivalence ratio of $\phi_a=0.3$ and mean mixture fraction of $Z_a=0.0197$ at conditions shown in Figure~\ref{FIG_FLAMELET_STRUCTURE}(a). The mixture-fraction distribution at the combustor exit is represented, to a first approximation, by a beta-distribution, $\beta(z)$. The fluctuation magnitude is estimated as $\delta Z_a=\sqrt{\zeta Z_a(1-Z_a)},$ where $\zeta\in[0,1]$ is a coefficient for the mixedness~\citep{DIMOTAKIS_MILLER_POF1990}. In a combustor in which the mixing is nearly completed with $\zeta=10^{-4}$, the temperature fluctuation can be evaluated from $\delta T_a = \{\int^1_0 [T(z)-T_a]^2\beta(z)dz\}^{1/2}$, where $T(z)$ is the flame solution from Figure~\ref{FIG_FLAMELET_STRUCTURE}(a). Hence, one finds that $\xi_a/\sigma_a=0.015$, indicating that the noise ratio at subsonic condition is below 0.1. However, this ratio increases to values of 0.5 (supercritical nozzle) and exceeds values of 5 (supercritical nozzle with shock), as shown by the red dashed lines in Figure~\ref{FIG_TRANSFER_PLOTS}. This suggests that the compositional noise can become a relevant contributor to indirect combustion noise at these conditions. 

\subsection{Discussion}
The present analysis employed the compact-nozzle theory, which relies on simplifications that are not strictly valid for $\He>0$ (see \S\ref{SEC_NOISE_CONFINEMENT_MATH}). Therefore, the compact-nozzle assumption can be relaxed to consider effects of finite nozzle-length and wave phase differences, as discussed in \S\ref{SEC_INTRO}. In addition, multidimensional high-fidelity numerical simulations provide further opportunities to assess the importance of compositional noise, which is the subject of ongoing research.

This analysis also stipulates the need for experimental investigations to measure compositional noise and obtain a firm evaluation of the level of compositional inhomogeneities at the combustor exit in gas-turbine engines. Since the product species of \ce{CO2}, \ce{H2O}, and \ce{CO} are leading contributors to the chemical potential function, gas-sampling probes, tunable diode laser absorption spectroscopy, and other intrusive and non-intrusive techniques, could be employed to quantify the spatial and temporal evolution of compositional inhomogeneities. 
\section{Conclusions}
By modelling inhomogeneities in the gas composition exiting the combustor and entering a nozzle, the compositional noise is identified as a source of indirect combustion noise. To describe this source mechanism, the compact-nozzle theory is extended to consider a multi-component gas mixture and the chemical potential function. This theory is applied to subcritical and supercritical nozzle flows. It is found that the compositional noise exhibits strong dependence on the mixture composition, and can become comparable to -- and even exceed -- direct noise and entropy noise for supercritical nozzles and lean mixtures. This suggests that compositional noise may require consideration with the implementation of low-emission combustors, high power-density engine cores, or compact burner concepts~\citep{HULTGREN_CORENOISE_WG2011,CHANG_LEE_HERBON_KRAMER_JAAE2013}.

\section*{Acknowledgments}
Financial support through NASA with award number NNX15AV04A and the Ford-Stanford Alliance project \#C2015-0590 is gratefully acknowledged. The authors are grateful to Dr. Lucas Esclapez for his help with the flamelet calculations. 
\bibliographystyle{jfm}

\begin{thebibliography}{29}
\expandafter\ifx\csname natexlab\endcsname\relax\def\natexlab#1{#1}\fi
\def\au#1{#1} \def\ed#1{#1} \def\yr#1{#1}\def\at#1{#1}\def\jt#1{\textit{#1}}
  \def\bt#1{#1}\def\bvol#1{\textbf{#1}} \def\vol#1{#1} \def\pg#1{#1}
  \def\publ#1{#1}\def\arxiv#1{#1}\def\org#1{#1}\def\st#1{\textit{#1}}

\bibitem[Bake {\em et~al.\/}(2009)Bake, Richter, M\"uhlbauer, Kings, R\"ohle,
  Thiele \& Noll]{BAKE_RICHTER_MUEHLBAUER_KINGS_ROEHLE_THIELE_NOLL_JSV2009}
{\sc \au{Bake, F.}, \au{Richter, C.}, \au{M\"uhlbauer, C.}, \au{Kings, N.},
  \au{R\"ohle, I.}, \au{Thiele, F.} \& \au{Noll, B.}} \yr{2009}  \at{The
  entropy wave generator {(EWG)}: A reference case on entropy noise}.  \jt{J.
  Sound Vib.}  \bvol{326},  \pg{574--598}.

\bibitem[Candel {\em et~al.\/}(2009)Candel, Durox, Ducruix, Birbaud, Noiray \&
  Schuller]{CANDEL_DUROX_DUCRUIX_BIRBAUD_NOIRAY_SCHULLER_IJAA2009}
{\sc \au{Candel, S.}, \au{Durox, D.}, \au{Ducruix, S.}, \au{Birbaud, A.-L.},
  \au{Noiray, N.} \& \au{Schuller, T.}} \yr{2009}  \at{Flame dynamics and
  combustion noise: {Progress} and challenges}.  \jt{Int. J. Aeroacoustics}
  \bvol{8}~(1-2),  \pg{1--56}.

\bibitem[Candel(1972)]{CANDEL_DISS1972}
{\sc \au{Candel, S.~M.}} \yr{1972}  \at{Analytical studies of some acoustic
  problems of jet engines}. PhD thesis, California Institute of Technology.

\bibitem[Chang {\em et~al.\/}(2013)Chang, Lee, Herbon \&
  Kramer]{CHANG_LEE_HERBON_KRAMER_JAAE2013}
{\sc \au{Chang, C.~T.}, \au{Lee, C.-M.}, \au{Herbon, J.~T.} \& \au{Kramer,
  S.~K.}} \yr{2013}  \at{{NASA} environmentally responsible aviation project
  develops next-generation low-emissions combustor technologies ({Phase I})}.
  \jt{J. Aeronaut. Aerospace Eng.}  \bvol{2}~(4),  \pg{1000116}.

\bibitem[Chu \& Kov{\'{a}}sznay(1958)]{CHU_KOVASZNAY_JFM1958}
{\sc \au{Chu, B.~T.} \& \au{Kov{\'{a}}sznay, L. S.~G.}} \yr{1958}
  \at{{Non-linear interactions in a viscous heat-conducting compressible gas}}.
   \jt{J. Fluid Mech.}  \bvol{3},  \pg{494--514}.

\bibitem[Cumpsty(1979)]{Cumpsty1979}
{\sc \au{Cumpsty, N.~A.}} \yr{1979}  \at{{Jet engine combustion noise:
  Pressure, entropy and vorticity perturbations produced by unsteady combustion
  or heat addition}}.  \jt{J. Sound Vib.}  \bvol{66}~(4),  \pg{527--544}.

\bibitem[Dimotakis \& Miller(1990)]{DIMOTAKIS_MILLER_POF1990}
{\sc \au{Dimotakis, P.~E.} \& \au{Miller, P.~L.}} \yr{1990}  \at{Some
  consequences of the boundedness of scalar fluctuations}.  \jt{Phys. Fluids}
  \bvol{2}~(11),  \pg{1919--1920}.

\bibitem[Dowling \& Mahmoudi(2015)]{DOWLING_MAHMOUDI_PCI2015}
{\sc \au{Dowling, A.~P.} \& \au{Mahmoudi, Y.}} \yr{2015}  \at{Combustion
  noise}.  \jt{Proc. Combust. Inst.}  \bvol{35},  \pg{65--100}.

\bibitem[Duran \& Moreau(2013)]{DURAN_MOREAU_JFM2013}
{\sc \au{Duran, I.} \& \au{Moreau, S.}} \yr{2013}  \at{Solution of the
  quasi-one-dimensional linearized {Euler} equations using flow invariants and
  the {Magnus} expansion}.  \jt{J. Fluid Mech.}  \bvol{723},  \pg{190--231}.

\bibitem[Giauque {\em et~al.\/}(2012)Giauque, Huet \&
  Clero]{GIAUQUE_HUET_CLERO_JEGTP2012}
{\sc \au{Giauque, A.}, \au{Huet, M.} \& \au{Clero, F.}} \yr{2012}
  \at{Analytical analysis of indirect combustion noise in subcritical nozzles}.
   \jt{J. Eng. Gas Turbines Power}  \bvol{134}~(111202),  \pg{1--8}.

\bibitem[Goh \& Morgans(2011)]{GOH_MORGANS_JSV2011}
{\sc \au{Goh, C.~S.} \& \au{Morgans, A.~S.}} \yr{2011}  \at{Phase prediction of
  the response of choked nozzles to entropy and acoustic disturbances}.  \jt{J.
  Sound Vib.}  \bvol{330},  \pg{5184--5198}.

\bibitem[Goodwin(1998)]{CANTERA}
{\sc \au{Goodwin, D.~G.}} \yr{1998} {\sc{Cantera:}} {An} open-source,
  object-oriented software suite for combustion.

\bibitem[Hultgren(2011)]{HULTGREN_CORENOISE_WG2011}
{\sc \au{Hultgren, L.~S.}} \yr{2011} Core noise: Implications of emerging {N+3}
  designs and acoustic technology needs. Acoustics Technical Working Group.

\bibitem[Hurle {\em et~al.\/}(1968)Hurle, Price, Sugden \&
  Thomas]{HURLE_PRICE_SUGDEN_THOMAS_PRSLA1968}
{\sc \au{Hurle, I.~R.}, \au{Price, R.~B.}, \au{Sugden, T.~M.} \& \au{Thomas,
  A.}} \yr{1968}  \at{Sound emission from open turbulent premixed flames}.
  \jt{Proc. R. Soc. London A}  \bvol{303},  \pg{409--427}.

\bibitem[Ihme(2017)]{IHME_ARFM2017}
{\sc \au{Ihme, M.}} \yr{2017}  \at{Combustion and engine-core noise}.
  \jt{Annu. Rev. Fluid Mech.} In press.

\bibitem[Ihme {\em et~al.\/}(2009)Ihme, Pitsch \&
  Bodony]{IHME_PITSCH_BODONY_PCI32}
{\sc \au{Ihme, M.}, \au{Pitsch, H.} \& \au{Bodony, D.}} \yr{2009}
  \at{Radiation of noise in turbulent non-premixed flames}.  \jt{Proc. Combust.
  Inst.}  \bvol{32},  \pg{1545--1553}.

\bibitem[Job \& Herrmann(2006)]{JOB_HERRMANN_EJP2006}
{\sc \au{Job, G.} \& \au{Herrmann, F.}} \yr{2006}  \at{Chemical potential -- a
  quantity in search of recognition}.  \jt{Eur. J. Phys.}  \bvol{27}~(2),
  \pg{353--371}.

\bibitem[Keck \& Gillespie(1971)]{KECK_GILLESPIE_CF1971}
{\sc \au{Keck, J.~C.} \& \au{Gillespie, D.}} \yr{1971}  \at{Rate-controlled
  partial-equilibrium method for treating reacting gas mixtures}.  \jt{Combust.
  Flame}  \bvol{17},  \pg{237--241}.

\bibitem[Kings \& Bake(2010)]{KINGS_BAKE_IJSCD2010}
{\sc \au{Kings, N.} \& \au{Bake, F.}} \yr{2010}  \at{Indirect combustion noise:
  {Noise} generation by accelerated vorticity in a nozzle flow}.  \jt{Int. J.
  Spray Combust. Dyn.}  \bvol{2}~(3),  \pg{253--266}.

\bibitem[Marble \& Candel(1977)]{MARBLE_CANDEL_JSV1977}
{\sc \au{Marble, F.~E.} \& \au{Candel, S.~M.}} \yr{1977}  \at{Acoustic
  disturbance from gas non-uniformities convected through a nozzle}.  \jt{J.
  Sound Vib.}  \bvol{55}~(2),  \pg{225--243}.

\bibitem[Moase {\em et~al.\/}(2007)Moase, Brear \&
  Manzie]{MOASE_BREAR_MANZIE_JFM2007}
{\sc \au{Moase, W.~H.}, \au{Brear, M.~J.} \& \au{Manzie, C.}} \yr{2007}
  \at{The forced response of choked nozzles and supersonic diffusers}.  \jt{J.
  Fluid Mech.}  \bvol{585},  \pg{281--304}.

\bibitem[Peters(2000)]{PETERS_BOOK2000}
{\sc \au{Peters, N.}} \yr{2000} {\em Turbulent Combustion\/}.  \publ{Cambridge:
  Cambridge University Press}.

\bibitem[Rajaram \& Lieuwen(2003)]{RAJARAM_LIEUWEN_CST2003}
{\sc \au{Rajaram, R.} \& \au{Lieuwen, T.}} \yr{2003}  \at{Parametric studies of
  acoustic radiation from premixed flames}.  \jt{Combust. Sci. Tech.}
  \bvol{175}~(12),  \pg{2269--2298}.

\bibitem[Singh {\em et~al.\/}(2005)Singh, Zhang, Gore, Mongeau \&
  Frankel]{SINGH_ZHANG_GORE_MONGEAU_FRANKEL_PCI30}
{\sc \au{Singh, K.~K.}, \au{Zhang, C.}, \au{Gore, J.~P.}, \au{Mongeau, L.} \&
  \au{Frankel, S.~H.}} \yr{2005}  \at{An experimental study of partially
  premixed flame sound}.  \jt{Proc. Combust. Inst.}  \bvol{30},
  \pg{1707--1715}.

\bibitem[Stow {\em et~al.\/}(2002)Stow, Dowling \&
  Hynes]{STOW_DOWLING_HYNES_JFM2002}
{\sc \au{Stow, S.~R.}, \au{Dowling, A.~P.} \& \au{Hynes, T.~P.}} \yr{2002}
  \at{Reflection of circumferential modes in a choked nozzle}.  \jt{J. Fluid
  Mech.}  \bvol{467},  \pg{215--239}.

\bibitem[Strahle(1978)]{STRAHLE_PECS1978}
{\sc \au{Strahle, W.~C.}} \yr{1978}  \at{Combustion noise}.  \jt{Prog. Energy
  Combust. Sci.}  \bvol{4},  \pg{157--176}.

\bibitem[Vie {\em et~al.\/}(2015)Vie, Franzelli, Gao, Lu, Wang \& Ihme]{Vie15}
{\sc \au{Vie, A.}, \au{Franzelli, B.}, \au{Gao, Y.}, \au{Lu, T.}, \au{Wang, H.}
  \& \au{Ihme, M.}} \yr{2015}  \at{Analysis of segregation and bifurcation in
  turbulent spray flames: {A 3D} counterflow configuration}.  \jt{Proc.
  Combust. Inst.}  \bvol{35}~(2),  \pg{1675--1683}.

\bibitem[Williams(1985)]{WILLIAMS_BOOK1985}
{\sc \au{Williams, F.~A.}} \yr{1985} {\em Combustion Theory\/}.  \publ{Reading,
  MA: Perseus Books}.

\bibitem[Zhao \& Frankel(2001)]{ZHAO_FRANKEL_POF2001}
{\sc \au{Zhao, W.} \& \au{Frankel, S.~H.}} \yr{2001}  \at{Numerical simulations
  of sound radiated from an axisymmetric premixed reacting jet}.  \jt{Phys.
  Fluids}  \bvol{13}~(9),  \pg{2671--2681}.

\end{thebibliography}

\end{document}